\documentstyle[12pt]{article}
  \advance\oddsidemargin  by -1in
  \advance\evensidemargin by -1in
\def\lsim{\mathrel{\rlap{\lower4pt\hbox{\hskip1pt$\sim$}}
    \raise1pt\hbox{$<$}}}         %less than or approx. symbol
\def\gsim{\mathrel{\rlap{\lower4pt\hbox{\hskip1pt$\sim$}}
    \raise1pt\hbox{$>$}}}         %greater than or approx. symbol

\def\be{\begin{equation}}
\def\ee{\end{equation}}
\def\bq{\begin{eqnarray}}
\def\eq{\end{eqnarray}}

\begin{document}

\pagestyle{empty}

\begin{flushright}
DFTT 24/95\\
April 1995
\end{flushright}

\vspace{2 cm}

\begin{center}
{\bf \large THE STRANGE QUARK DISTRIBUTION\\}

\vspace{1 cm}

V.~BARONE$^{a,}$\footnote{Also at II Facolt{\`a} di Scienze MFN, 15100
Alessandria, Italy.}, M.~GENOVESE$^{a,b}$,
N.N.~NIKOLAEV$^{c,d}$,\\
E.~PREDAZZI$^{a}$ and B.G.~ZAKHAROV$^{d}$\\

\vspace{1 cm}

{\sl $^{a}$Dipartimento di Fisica Teorica dell'Universit{\`a}\\
and INFN, Sezione di Torino, I--10125 Torino, Italy\smallskip\\
$^{b}$Theory Division,
CERN, CH--1211 Gen{\`e}ve 23, Switzerland \smallskip \\
$^{c}$Institut f{\"u}r Kernphysik, Forschungszentrum J\"ulich,\\ D-52425
J\"ulich, Germany\smallskip\\
$^{d}$L.~D.~Landau Institute for
Theoretical Physics,\\ GSP-1, 117940 Moscow V-334, Russia
\vspace{1.2cm}\\ }
\underline{ABSTRACT}\medskip\\
\end{center}

We discuss the latest CCFR determination of the strange sea density
of the proton. We comment on the differences with a previous,
leading--order, result and point out the relevance of
quark mass effects and current non--conservation effects. By
taking
them into account it is possible
to solve the residual discrepancy with
another determination of the strange quark
distribution. Two important sources
of uncertainties are also analyzed.

\vfill
\eject

\pagestyle{plain}
\baselineskip 24pt

\section{Introduction}
\bigskip

Until few
years ago the problem of the strange sea distribution
was rather controversial,
despite the impressive amount of knowledge on the
internal structure of nucleons
accumulated in the last decade. The best available fits in 1993
\cite{MRS1,CTEQ} differed by almost a factor 2  for the strange
density. As we pointed out in \cite{BGNPZ3,BGNPZ4},
this puzzling situation was mainly determined by an
erroneous interpretation of the experimental results,
which ignored important physical effects already
investigated in \cite{BGNPZ1,BGNPZ2}.

There are two ways to extract the strange sea distribution $s(x)$.
The first method consists in subtracting the structure functions
$F_2$ measured in muon ($\mu$) and neutrino
($\nu$) deep inelastic scattering (DIS). We call this
the $\nu- \mu$ determination of $s(x)$. The second, more direct,
 method
consists in selecting $\nu$DIS events
with charm production: the signature of these events is
the presence of opposite--sign dimuons in the final state.

When the NMC $\mu$DIS data \cite{NMC} and the CCFR $\nu$DIS
data \cite{CCFR3,CCFR1} made both these determinations possible, it
was found, rather surprisingly, that the two results for $s(x)$ were
largely different (see however the anticipations
in \cite{BGNPZ1,BGNPZ2}): the strange density extracted from dimuon
data was considerably smaller than the one obtained by
subtracting $\nu$DIS and $\mu$DIS data.

In trying to solve this discrepancy all the available
global parton parametrizations ran into serious difficulties.
The CTEQ group, trusting
the $\nu - \mu$ result, was led to advocate a very large strange
sea content $\kappa = 2S/ (\bar U + \bar D) \simeq 0.9 \,\, \,
(S \equiv \int {\rm d} x \, x s(x)$, etc.), {\it i.e.}  an
almost $SU(3)$--symmetric light sea. At the same time, however,
the CTEQ
strange distribution
lied well above the dimuon data.
On the other hand, the MRS $D_0'$ fit stuck to the more
conservative value $\kappa = 0.5$ (which was an input) but still
overshot the dimuon data, while lying below the $\nu-\mu$ result.
(We shall see {\it a posteriori} that the MRS
compromise between the two data sets and the  value of $\kappa
\simeq 0.5$ are --accidentally-- much closer to the truth than the
CTEQ picture.)

The solution to the strange sea puzzle proposed in
\cite{BGNPZ3,BGNPZ4} was very simple: there is no real puzzle
because the two determinations of $s(x)$ measure
in fact different quantities, none of which coincides with the
true strange density.
This is due to the fact that, up to moderately large $Q^2$ (of order
of $30 \, GeV^2$ or so, {\it i.e.}
not much above the charm threshold),
the relevant diagrams for charm production
are the vector--boson--gluon fusion subprocesses \cite{GR}. These
are, in the common massless QCD terminology, {\it
next--to--leading} order diagrams and hence they are often mistakenly
forgotten as if they were {\it subleading} corrections.
When the gluon--fusion diagrams are taken into account,
two
effects arise which
explain why the two determinations of $s(x)$ do not really
provide $s(x)$ (at least directly, as a naive
leading--order analysis would suggest)
and should indeed give different results. They are:
{\it i)} quark mass corrections;
{\it ii)}
non conservation of weak currents (yielding large longitudinal
contributions).
These effects are calculable in perturbative QCD
\cite{BGNPZ1,BGNPZ2} and are relevant up to moderate
$Q^2$ values
(not much larger than the heavy quark mass scale). We called
them {\it non--universality} effects because they make
the heavy flavour contributions to $\nu$DIS and
to $\mu$DIS structure functions {\it intrinsically} different.

Near threshold, massless QCD is inappropriate to describe
heavy quark production and mass contributions must be kept.
The gluon--fusion process is the dominant one. Since
the mass thresholds in the transitions
$W^{+}\bar s \rightarrow c$, $\gamma^{*}
\rightarrow \bar ss(\bar c c)$, $Z^{0}
\rightarrow \bar s s (\bar c c)$,
are different,
and since the longitudinal structure functions in weak DIS
are larger than in electromagnetic DIS because weak
currents are non conserved, we expect \cite{BGNPZ1,BGNPZ2}
\be
F_2^{\nu, \bar s c} \neq F_2^{\nu, \bar s s} \neq F_2^{\nu, \bar c c}
\neq F_2^{\mu, \bar s s} \neq F_2^{\mu, \bar c c}\,\,,
\label{i1}
\ee
where we denote by $F_2^{\nu, \bar s c}$ the $\bar s c$ contribution
to the $\nu$DIS structure function $F_2$ with the electroweak
coupling factored out
(differently stated, $F_2^{\nu, \bar s c}$ reduces to $x(\bar s + c)$
at leading order),
and we use an analogous notation for the other quantities.

Due to the importance of the gluon--fusion processes,
we must expect a considerable difference between a
leading order (LO)
 and a next--to--leading (NLO)
extraction of $s(x)$ from the dimuon experiment
\cite{BGNPZ3,BGNPZ4}.
This expectation proves correct. The CCFR collaboration
has recently released \cite{CCFR2} a new determination of $s(x)$ based
on an analysis of the dimuon data which takes into account
the gluon--fusion processes. The `new' strange density
\cite{CCFR2} is considerably larger than the `old' one
\cite{CCFR1} and partially bridges up the gap with the
$\nu - \mu$ result.

The purpose of this paper is to discuss and clarify our
present knowledge of the strange density.
We shall comment on the recent CCFR determination of $s(x)$,
on its difference with the previous one, and on the
important physics behind such a difference.
On the quantitative side, we shall show that the residual
discrepancy existing at small $x$ between the CCFR NLO
strange density and the $\nu-\mu$ data is easily
accounted for by the non--universality (or, differently
said, by next-to-leading order) effects related to the gluon fusion
processes. Practically, our calculations endorse an
MRS--like strange sea density \cite{MRS2}
with a value $\kappa \simeq 0.5$.
We shall also
discuss some subtleties and some sources of uncertainties
in the analysis of the neutrino data, and propose
a more convenient way to present the dimuon results.

\vspace{1.5cm}
\section{The strange density from dimuon data}
\bigskip

The most direct mechanism to measure the strange density is
charm production in charged current neutrino deep inelastic scattering.

The $\nu$DIS cross section reads \cite{libri}
\be
\frac{{\rm d}^2 \sigma^{\nu N}}{{\rm d}x {\rm d}y}
= \frac{G^2 m_N E_{\nu}}{\pi (1 + \frac{Q^2}{M_W^2})^2}
\, \left [ x y^2 F_1^{\nu N}(x) + (1-y) F_2^{\nu N}(x) +
(y - \frac{y^2}{2}) x F_3^{\nu N}(x) \right ] \,\,.
\label{a1}
\ee
If we restrict ourselves to charm excitation,
only the transitions $ W^{+}d \rightarrow c, \, W^{+}s
\rightarrow c$ contribute to the structure functions.
At leading order we have to consider only the quark
scattering terms \footnote{For the sake of clarity
we stick hereafter to the terminology
of Ref.~\cite{Tung}, which is used by the CCFR
collaboration in their analysis.} and the
cross section for charm production in $\nu$DIS can be
expressed in terms of the LO parton densities as
\be
\frac{{\rm d}^2
\sigma(\nu N \rightarrow c X)}{{\rm d}x {\rm d}y}
= \frac{G^2 m_N E_{\nu}}{\pi (1 + \frac{Q^2}{M_W^2})^2}
\, x \left \{ \left [ u(x,Q^2) + d(x,Q^2) \right ]
\vert V_{cd} \vert^2 + 2 s(x,Q^2) \vert V_{cs} \vert^2 \right \}
\label{a2}
\ee
where $V_{cd}, V_{cs}$ are the Cabibbo--Kobayashi--Maskawa
matrix elements. Since $V_{cs} \gg V_{cd}$ the measure
of $\sigma(\nu N \rightarrow c X)$ provides an excellent
determination of $xs(x)$.

In order to take into account effects connected with
the non negligible mass of the charm quark, it has been
customary for many years to adopt the slow--rescaling procedure
(this is what the CCFR collaboration also did in \cite{CCFR1}).

In practice, the slow rescaling method consists in replacing
Bjorken's $x$ by the new variable
$\xi = x (1 + m_c^2/Q^2)$, which is (naively) expected
to take into account the effects of the
heavy quark mass.
If the Callan--Gross relation is enforced
in terms of $\xi$, namely $F_2(\xi) = 2 \xi F_1$,
 an extra factor appears in the
$\nu$DIS cross section  which then reads
\bq
\frac{{\rm d}^2
\sigma(\nu N \rightarrow c X)}{{\rm d}\xi {\rm d}y}
&=& \frac{G^2 m_N E_{\nu}}{\pi (1 + \frac{Q^2}{M_W^2})^2}
\, \nonumber  \\
&\times&
\xi \left \{ \left [ u(\xi,Q^2) + d(\xi,Q^2) \right ]
\vert V_{cd} \vert^2 + 2 s(\xi,Q^2) \vert V_{cs} \vert^2 \right \}
\, \left ( 1 + \frac{m_c^2}{2 m_N E_{\nu} \xi} \right )\,\,.
\label{a2b}
\eq

Slow rescaling lacks a solid theoretical foundation. It is a
sensible method if one considers only the quark scattering term.
In this case, in the $W^{+}s \rightarrow c$ transition, the
$s$ quark is taken on shell, as usual in the parton model,
and its mass is neglected with respect to the charm mass:
if we call $\xi$ the fraction of proton's momentum carried
by tha strange quark, the
substitution $x \rightarrow \xi$ follows straightforwardly.

However, considering higher order, gluon fusion,
 diagrams (which are actually
the dominant ones) it is no longer  possible to assume that
the $s$ quark is on shell: the whole procedure thus breaks up
and a more sophisticated treatment is called for. Of course, one could
still think that slow rescaling mimicks rather faithfully
the real world and that it accounts for
quark mass effects in an {\it effective} way.
This attitude has been quite popular and it is still widely believed
that slow rescaling is at least a very good approximation.
One of the conclusions of the first, leading order, CCFR
analysis was that the data supported the slow rescaling model
of charm production (although with a very small charm mass,
$m_c \simeq 1.31 \, GeV^2$). The NLO analysis has completely
reversed the situation: a large difference is found between
the LO determination with slow rescaling and the new one,
which is on a better theoretical ground and leads to
a more realistic charm mass, $m_c \simeq 1.70 \, GeV^2$.
The first conclusion we can draw from the new results
is indeed that the slow rescaling, which is intrinsically
a LO procedure, fails to provide a realistic picture of
charm production and therefore
must simply be abandoned (this criticism
was anticipated in \cite{BGNPZ1},
well before  both CCFR determinations).

In the region of small and moderate values of
$Q^2$ (where most of the CCFR data lie),{\it i.e.} not much above
the charm threshold $m_c^2$, it is not legitimate to retain
only the quark scattering LO diagrams. As a matter of fact, near
threshold,
the whole contribution of a heavy quark  to structure
functions\footnote{Remember that, in charged current $\nu$DIS,
probing the $s$ quark
means at the same time exciting the $c$ quark.} is given
by the vector--boson--gluon fusion process (Fig.~1), which are
conventionally classified as a next--to--leading order term
(although the leading term is not the dominant one).

In the case at hand
we have for $F_2$, in a formal notation ($\otimes$ means
convolution) \cite{GR,BGNPZ1}
\be
F_2^{\bar s c} = (\frac{\alpha_s}{2 \pi})\, g \otimes C_2 (
W^{+}g \rightarrow \bar s c)\,,
\label{a3}
\ee
where $g(x,Q^2)$ is the gluon distribution and $C_2$
is the
unsubtracted Wilson coefficient, {\it i.e.} the full cross
section for the $W^{+} g \rightarrow \bar s c$ process,
which is made of the two inseparable diagrams shown in Fig.~1.

It is worthwhile to draw a comparison with another approach
\cite{Tung}. In the formalism of Ref.~\cite{Tung}
the $\bar s c$ contribution to structure functions
is expressed as (we omit all the electroweak
couplings)
\be
F_2^{\bar s c} = x(\bar s + c) - C_2^{\rm subtr} +
(\frac{\alpha_s}{2 \pi})\,  g \otimes C_2 (
W^{+}g \rightarrow \bar s c)\, ,
\label{a4}
\ee
where $\bar s(x,Q^2)$ and $c(x,Q^2)$ are NLO parton densities, and
$C_2^{\rm subtr}$ is a subtraction term proportional to
$\alpha_s \log({\mu_f^2/m_c^2})$ ($\mu_f$ is the factorization scale).
In this approach the logarithmic term in the Wilson
coefficient, which explodes at
$Q^2$ much larger than $m_c^2$, giving eventually rise to
collinear singularities, is subtracted out; at the same
time, the quark scattering term is introduced so that,
when the physical scale is very large compared to the
heavy quark mass,
the massless QCD parton model is regained.

Up to $Q^2$ of order $10\, m_c^2$  the two formulas
for $F_2^{\bar s c}$ given above, eqs.~(\ref{a3}) and
(\ref{a4}), are equivalent \cite{Tung}.
In fact, the two extra--terms
in eq.~(\ref{a4}) are approximately equal and cancel out.
% besides,
%they are individually small compared to the $W$--gluon fusion
%contribution.
All the relevant physics (mass effects and
current non conservation effects producing large
longitudinal structure functions) is thus contained
in the $W$--gluon fusion diagrams.

When these are taken into account the relation between
the $\nu$DIS cross section with charm production and the
parton densities is much more involved than eqs.~(\ref{a2})
or (\ref{a2b}). In fact, the simplicity of the leading order
formulas for ${\rm d}^2
\sigma(\nu N \rightarrow c X)/{\rm d}\xi {\rm d}y$
is determined by the (fortuitously) virtuous combination
of two elements: {\it i)} the relation (valid
for the structure function components relevant to charm excitation)
\be
F_2^{\nu N} = 2 x F_1^{\nu N} = x F_3^{\nu N} =
x(u + d + 2s)\,,
\label{a5}
\ee
(or the corresponding one in terms of $\xi$), and {\it ii)}
the consequent, accidental, cancellation of all factors
in $y$ in front of the structure functions.
None of these two circumstances occur in the next--to--leading order
case (we keep using this terminology
%for a
%better understanding of this matter,
although we stressed above that
the $Wg$ fusion contribution is the {\it dominant} one
%--
%as a matter of fact, the {\it only}
%one in the Gl{\"u}ck--Reya scheme --
and cannot be, strictly
speaking, called {\it next--to--leading}).
The expression on the r.h.s. of eq.~(\ref{a2b}) is
therefore much more complicated and, of course,
the slow rescaling substitution becomes completely meaningless.

Hence it is expected, from a
theoretical viewpoint,
that there should be a large difference between the
leading order strange density determined from
(\ref{a4}) and the next--to--leading order strange
density extracted from (\ref{a1}) upon use of eq.~(\ref{a4}).
This is indeed what has been recently found \cite{CCFR2}
in the new, NLO,
CCFR analysis of the dimuon data. As one can see in Fig.~2,
at moderate $Q^2$ values,
the difference between the LO result for $xs(x)$ and the NLO result
is rather large (more than $50 \%$ at $Q^2 \simeq 10\, GeV^2$).
This is clearly not a mere higher--order correction: the
LO analysis hides and neglects all the important physics
contained in the $W$--gluon fusion diagrams. The slow rescaling
method, while theoretically
ill--founded, is not even an effectively
successful way to account
for heavy quark masses. Moreover, mass effects
are not the whole story. Effects of current non conservation in
$\nu$DIS scattering are quantitatively
as much (or even more) important \cite{BGNPZ2}. In $\nu$DIS
the ratio
$R = \sigma_L/\sigma_T$ in the heavy quark sector near threshold
 is larger than 1. Moreover it is
different
in charged current processes (where both the vector and the axial
currents are not conserved) and in  neutral current processes
(where only the axial current is not conserved). The
large longitudinal
contribution leads to a strong violation of the Callan--Gross
relation. Neither the use of this relation
nor its correction
by a value of $R$ taken from electromagnetic scattering (as it was
done in the LO analysis
\cite{CCFR1})
are then
legitimate. The inclusion of gluon fusion diagrams allows
considering all important physical effects automatically and in
a QCD computable way.

The new CCFR strange quark distribution is in much better agreement with
the MRS--A fit \cite{MRS2} (see Fig.~2)
and leads to a strange sea content
$\kappa = 0.477 \pm 0.050$ at the average $\langle
Q^2 \rangle \simeq 22\, GeV^2$.
There is a $(10-15)\%$ error on the result due to the factorization
scale uncertainty (see sec.~4). The measured quark mass is
$m_c = 1.70 \pm 0.19\, GeV^2$ and is larger than the unrealistic
value $m_c = 1.31\, GeV^2$ found in the LO analysis of
Ref.~\cite{CCFR1} because of the slow rescaling method adopted there.

Finally,
notice that,  in passing from LO to NLO, a variation
is expected (and detected \cite{CCFR3})
also for the non strange distributions, which
are Cabibbo suppressed in the cross section \ref{a2},
because the $u$ and
$d$ contributions to structure functions are mixed with
the charm contribution in the $W^{+}g \rightarrow \bar u c, \bar
d c $ processes.

\vspace{1.5cm}
\section{Comparison with another determination of the strange density}
\bigskip

There is another way to extract the strange sea density from deep
inelastic scattering. It combines measurements of $\nu$DIS
and $\mu$DIS structure functions:
we shall call it the $\nu - \mu$ determination
of $s(x)$.

Let us resort once more to the parton model or,
equivalently,
to leading order QCD. The decompositions of $F_2$ for
muonic and neutrino probes and an isoscalar
nucleon are
\bq
F_2^{\mu N} &=& \frac{5}{18} x (u + \bar u + d + \bar d)+
\frac{8}{9} x c + \frac{2}{9} x s\,\,,
\label{b1} \\
F_2^{\nu N} &=& x (u + \bar u + d + \bar d + 2s + 2 \bar c)\,.
\label{b2}
\eq
In eq.~(\ref{b1}) we used $s=\bar s, \, c= \bar c$ and eq.~(\ref{b2})
refers to charged current scattering.
By appropriately combining $F_2^{\nu N}$ and $F_2^{\mu N}$ one can select
the (leading order) strange distribution, under the
assumption $c \ll s$:
\be
\frac{5}{6} F_2^{\nu N}(x,Q^2) - 3 F_2^{\mu N}(x,Q^2)
= xs(x,Q^2)\,.
\label{b3}
\ee

It is immediately evident what is the main experimental difficulty
with this determination. The quantity on the l.h.s. of
eq.(\ref{b3}) is obtained by subtracting data from two
different experiments and is very sensitive to the relative
normalization. Besides, data are not taken at the same
$x$ and $Q^2$ values. Large uncertainties thus arise
in the $\nu - \mu$ difference.

On the other hand, the $\nu - \mu$ determination presents at least
two advantages. First of all, on the theoretical side, its
parton density content is simpler and much easier to
reconstruct than that of the quantity in curly brackets in
eq.~(\ref{a1}). Second, when $F_2^{\nu N}$ is measured
there is no spurious, acceptance--dependent, separation of the
two gluon fusion diagrams in Fig.~1, as in the dimuon
measurement (see next section), and the whole $\bar s c$
contribution to the structure function is determined.

Now, after the first (LO) CCFR determination of the
strange density came out \cite{CCFR1} it was clear that there existed
a big discrepancy (see Fig.~2) with the result from
the $\nu-\mu$ difference, obtained by
combining NMC \cite{NMC} and CCFR \cite{CCFR3} data, although the latter
was affected by large errors.
As we have already recalled, this
discrepancy was mainly due to the leading order analysis of
the dimuon data which hided important physical effects.
With the new CCFR determination of $xs(x)$ which includes
the contribution of the gluon fusion diagrams the situation
is considerably less puzzling. Still, there seems to be
a residual discrepancy  with the $\nu=\mu$ result, which deserves some
explanation (see Fig.~2).
This is simple if one remembers that the non--universality
effects taken into account in the NLO dimuon result are not included
in the LO formula (\ref{b3}) on which the identification of the
difference $(5/6) \, F_2^{\nu N} - 3 \, F_2^{\mu N}$ with
the strange density is based. Otherwise stated, NLO effects invalidate
eq.~(\ref{b3}) which must be replaced by
\be
\frac{5}{6} \, F_2^{\nu N} - 3 \, F_2^{\mu N} =
\frac{5}{3} \, F_2^{\bar s c} - \frac{1}{3} \, F_2^{s \bar s}\,\,,
\label{b4}
\ee
where (we are interested in the $Q^2$ region around $10\, GeV^2$)
\bq
F_2^{\bar s c} &=&
(\frac{\alpha_s}{2 \pi})\, g \otimes C_2 (
W^{+}g \rightarrow \bar s c) \,,
\label{b5} \\
F_2^{s \bar s} &=& x(s + \bar s)  +
(\frac{\alpha_s}{2 \pi})\, g \otimes C_2^{massless} (
\gamma^{*}g \rightarrow s \bar s )
 \simeq  x (s + \bar s) \,\,.
\label{b6}
\eq
The second equality in eq.~(\ref{b6}) is valid when the
physical scale $Q^2$ is sufficiently higher than the strange
threshold.

Were the non--universality ratio
$F_2^{\bar s c}/F_2^{s \bar s}$ equal to $1/2$,
as in the LO case, eq.~(\ref{b4}) would reduce to eq.~(\ref{b3}).
However, for not too large $Q^2$, this ratio is largely different
from $1/2$ (see \cite{BGNPZ3,BGNPZ4} and, for a
more systematic study, \cite{BDG}).
Therefore the difference $(5/6) F_2^{\nu N} - 3 F_2^{\mu N}$ does not
coincide with the strange quark distribution $x s(x)$.
We evaluated the r.h.s. of eq.~(\ref{b4}) by resorting to
eqs.~(\ref{b5},\ref{b6}), with factorization scale
$\mu_f^2 = m_c^2$,
 and using the MRS--A parton densities
\cite{MRS2}. Since the MRS--A strange density reproduces rather well
the CCFR NLO data, our calculation will also clarify whether
there is a real contradiction between the dimuon and the
$\nu-\mu$ determinations.
 The result
for the $\nu - \mu$ difference (\ref{b4})
is displayed in Fig.~3 and
compared to the data.
The good agreement found shows that
the dimuon and the $\nu-\mu$ determinations
are compatible in the whole $x$ range,
and is a check of the goodness of the MRS--A
parametrization of $xs(x)$. Notice that,
accounting for NLO effects, the $\nu-\mu$ difference
turns out to be larger than $xs(x)$ (dashed line)
at low $x$ and smaller
at high $x$, with a crossover at $x \simeq 0.07$.

%Thus, it is an easy task to estimate the correction to be made on the
%$\nu - \mu$ result in order
%to extract the NLO strange density $xs(x)$.
%Let us first isolate $xs(x)$:
%\be
%\frac{5}{6} \, F_2^{\nu N} - 3 \, F_2^{\mu N} = xs(x) \,
%( \frac{10}{3} \,
%\frac{F_2^{\bar s c}}{F_2^{s \bar s}} - \frac{2}{3}) \, \,.
%\label{b7}
%\ee
% Using this result,
%we can estimate the strange density
%from the $\nu - \mu$ measurement.
%To evaluate the quantity
%$F_2^{\bar s c}/F_2^{s \bar s}$ we resorted to eqs.(\ref{b5,b6})
%and to the parton densities parametrized by MRS \cite{MRS},
%which do not include the $\nu-\mu$ data. For the factorization scale
%with
%$m_c= 1.7 \, GeV$ and $m_s = 500 MeV$
% but
%we also checked that similar results are obtained with a different
%choice (for instance $\mu^2=Q^2$).
%We also stress that what we present here is
%a genuine, model--independent,
%QCD prediction.

With the (important) {\it caveat} illustrated in the next section,
we have now a trustworthy picture of the strange density. If we
believe in the recent CCFR determination and we assume the
reliability of the $\nu - \mu$ data, no discrepancy whatsoever is
at present detected and all experimental determinations converge to
an $s(x)$ well fitted by the MRS curve with a conservative
value of $\kappa$ ($\simeq 0.5$).

\vspace{1.5cm}
\section{Uncertainties in the extraction of the strange density}
\bigskip

We have seen that the strange density recently
determined by the
CCFR Collaboration
has been obtained by
a next--to--leading order QCD analysis of the
dimuon data, which considers all the
relevant effects previously overlooked.
The result, however, is affected by a relatively large
inherent uncertainty (more than $10 \%$).

In this section we want to discuss
two important sources of systematic
uncertainties in the extraction
of the strange sea density: one is specific of the dimuon
determination, the other is more generally related to the
kinematical region considered, close to a heavy quark threshold.

The first correction was already discussed in
\cite{BGNPZ4},  where it was pointed out that
in the first CCFR extraction
of the strange density an important acceptance effect had been
neglected, namely the experimentally
different weight of the two diagrams in
Fig.~1 due to the energy cut on the second muon.

In order to limit the background (mostly due to
kaon and pion decays) a lower cutoff is set on
the momentum $p_{\mu_2}$
of the muon coming from the semileptonic decay of charm ($p_{\mu_2}
\gsim 5 \, GeV$). This reflects itself into a cut on
the momentum of the produced $c$ quark (we shall call
$z$ the fraction of the light--cone momentum of the $W$ boson
carried by the charm). Given that the low--$z$ region
is dominated by the subprocess where the gluon splits into
a $c \bar c$ pair with the $\bar s$ produced in the
$W$--hemisphere (u channel), and, {\it viceversa}, the high--$z$
region is dominated by the process where the gluon splits
into an $s \bar s$ pair with the $c$ produced in the $W$--hemisphere,
it is clear that the cut on $z$ introduces an acceptance correction
on the final result for the strange density.
A way to compute such correction is to look at
the $z$--distribution $\sigma_{Wg}(z)$
for the $W$--gluon fusion process \cite{BGNPZ4}. This is
strongly asymmetric: the two peaks at $z\rightarrow 0$ (backward
peak) and at $z \rightarrow 1$ (forward peak) are
not delta--like at small $Q^2$ (they
tend to become so only at asymptotically large $Q^2$) and
have a different
$Q^2$ evolution: the forward peak appears sooner.
The $p_{\mu_2}$ cut can be taken to generate
an effective cutoff $z_c$ on $z$.
The $W$--gluon fusion
contribution to the structure function
$F_2^{\bar s c}$ really measured is then proportional
to $\int_{z_c}^{1} \, \sigma_{Wg}(z) {\rm d} z$ and hence smaller than
the whole contribution $\int_{0}^{1} \, \sigma_{Wg}(z) {\rm d} z$.
Were $z_c$ known, the correction would be theoretically predictable.
On the other hand a possible
uncertainty on $z_c$ may be a non irrelevant  source of
error on the extracted strange structure function.
In Fig.~4 we show the results of the model of Ref.~\cite{BGNPZ4} for
$F_2^{\bar s c}$ with two choices for $z_c$: $z_c =0.5$ (solid curve)
and $z_c= 0.8$ (dashed curve). The area below these two curves amounts
to about
$75 \%$ and $55 \%$, respectively,
of the whole integral (corresponding
to $z_c =0$).

An experimental evaluation
of this acceptance correction can be performed by folding the
$z$--cross section $\sigma_{Wg}(z)$ with an empirically
determined acceptance function, which is zero at $z \rightarrow 0$
and rises to 1 at large $z$
(the method sketched in the previous paragraph
corresponds to a step--function choice for
the acceptance curve). In its latest
analysis \cite{CCFR2}, the CCFR collaboration has carried out such
computation and found that the acceptance
correction is $60 \%$ with an estimated error of $10 \%$. These
values correspond approximately, in our approach,
to the situation depicted in
Fig.~4 and
confirm
the importance of the effect (and also, incidentally, the educated guess
on the acceptance uncertainty made in \cite{BGNPZ4}).

A more relevant (and fundamental) uncertainty on the strange quark
density
comes from
the fact that at $Q^2$ not much below a heavy quark threshold it is
generally unsafe to extract the {\it distributions} of
the vector--boson--gluon fusion products, instead of their
contributions to {\it structure functions}.

To explain this point, let us look for instance at the expression
(\ref{a4}) for $F_2^{\bar s c}$. Since the $W$--quark fusion
terms are negligible
with respect to the $W$--gluon fusion ones,
the (anti)quark
distributions are contained only in the quark scattering term.
This is small near threshold and undergoes a subtle and rather
precise cancellation with the subtraction term: therefore its extraction
is a delicate matter. More importantly,
the quark scattering contribution is strongly dependent
on the factorization scale, as it is theoretically
predicted \cite{Tung} and experimentally observed \cite{CCFR2}:
most of the overall $(10 - 15) \%$ uncertainty in the NLO CCFR
result for $xs(x)$
comes from the arbitrariness in the choice of $\mu^2$. By contrast,
the structure function $F_2^{\bar s c}$ is rather stable against
various choices of the factorization scale and is therefore
the best quantity to explore, at least as far as the quark
scattering terms do not become the dominant contribution -- which
happens at $Q^2$ well above threshold.

It would thus be desirable, at least near heavy quark thresholds,
 to get from experiments data on {\it structure functions} instead
of data on {\it parton distributions}.

\vspace{1.5cm}
\section{Conclusions}
\bigskip

Let us summarize the main points of this work.

We have by now a much better knowledge of the strange content of
the nucleon, coming mostly from neutrino deep inelastic scattering.

The next--to--leading order determination of $xs(x)$
performed by CCFR supersedes the leading--order one:
the latter should be recorded as a result which has little to do
with the real world. It relied on two assumptions:
{\it i)} the gluon fusion diagrams are negligible corrections;
{\it ii)}
the quark mass effects are accounted for by slow rescaling.
In the $Q^2$ region of present experimental interest, we pointed out
that:
the former assumption is simply wrong (as the explicit calculations
show); the latter is ill--established and the
slow rescaling procedure does not even
mimick the correct treatment of heavy quarks.
Not far from threshold,
the {\it next--to--leading order} term is
{\it not} a correction of the {\it leading--order} term
and one should be very careful in using such a terminology
which could induce into dangerous misunderstandings.

The discrepancy between the dimuon and the $\nu-\mu$ results
for $xs(x)$, which was worrisome at the time of the first
CCFR determination, is now solved. The higher--order
analysis correctly takes into account the effects which were
the physical cause of such an apparent puzzle
(we dubbed them  ``non--universality effects''): different
mass thresholds in neutral and charged current DIS and
large longitudinal contributions to $\nu$DIS structure
functions. In other terms, the dimuon data and the
$\nu - \mu$ data measure different quantities, which
coincide only when the two classes
of effects just mentioned are neglected,
{\it i.e.} only when a LO analysis is performed.
%This
%was predicted many years ago \cite{xx} and subsequently proposed as
%a solution of the strange sea puzzle \cite{xx}.
By accounting for these
non--universality (or NLO) effects
({\it i.e.}
for the non negligible difference between $2 F_2^{\bar s c}$
and $F_2^{c\bar c} + F_2^{s \bar s}$), we have explicitly shown that
even the
residual gap with the new dimuon data at small $x$
is fictitious.

Although greatly improved, our present knowledge
of the strange distribution is not yet free from uncertainties.
The NLO extraction of the strange sea density from dimuon data
near charm threshold is intrinsically unsafe, because the
quark scattering term (that is, the parton density)
is a small contribution
subjected to cancellation by the subtraction term and, at the same
time,
has a relatively strong dependence on the factorization scale.
This dependence weakens if one considers
the whole structure function
(including the dominant gluon fusion term).
Thus, at least at moderate values
of $Q^2$, the strange and charm structure functions (and not their
parton distributions) should be experimentally extracted to
be
object of theoretical
study.

\pagebreak

\pagebreak

\begin{center}

{\large \bf Figure captions}

\end{center}

\vspace{1cm}

\begin{itemize}

\item[Fig.~1]
$W$--gluon fusion process for charm production: a) u channel diagram;
b) t channel diagram.

\item[Fig.~2]
The strange quark distribution $xs(x)$. The boxes are the LO
CCFR determination \cite{CCFR1}
 at $Q^2 \simeq 10 \,  GeV^2$. The circles are the quantity
$(5/6) \,
F_2^{\nu N} - 3 \, F_2^{\mu N}$
at $Q^2 \simeq 7\, GeV^2$,
obtained by combining
NMC \cite{NMC} and CCFR data \cite{CCFR3}.
The shaded area
represents the new NLO CCFR result \cite{CCFR2} at $Q^2 = 10\, GeV^2$.
The continuous line is the MRS--A fit \cite{MRS2} for
$xs(x)$ at $Q^2 = 7 \, GeV^2$.
%The diamonds are the central points of the
%$\nu- \mu$ data corrected for higher--order
%(non--universality) effects.

\item[Fig.~3]
The difference
$(5/6) \,
F_2^{\nu N} - 3 \, F_2^{\mu N}$
at $Q^2 \simeq 7\, GeV^2$. The circles are the experimental results
obtained by combining
NMC \cite{NMC} and CCFR data \cite{CCFR3}.
The solid line is the next--to--leading order
QCD prediction described in the text.
The dashed line is the MRS--A fit \cite{MRS2} for
$xs(x)$ at $Q^2 = 7 \, GeV^2$.

\item[Fig.~4]
The charm--strange structure function $F_2^{\bar s c}$ at $Q^2 =
10\, GeV^2$
with two
different values of the cutoff $z_c$ on the light--cone momentum
of the charmed quark (see text): the solid line is for $z_c = 0.5$,
the dot--dashed line for $z_c = 0.8$.

\end{itemize}


\begin{thebibliography}{99}

\bibitem{MRS1}
A.D.~Martin, W.J.~Stirling and R.G.~Roberts, Phys. Lett.
 {\bf B306} (1993) 145; J. Phys. {\bf G19} (1993) 1429.

\bibitem{CTEQ}
J.~Botts {\it et al.} (CTEQ), Phys. Lett. {\bf B304} (1993) 159.

\bibitem{BGNPZ3}
V.~Barone, M.~Genovese,
N.N.~Nikolaev, E.~Predazzi and B.G.~Zakharov,
Phys. Lett. {\bf B317} (1993) 433.

\bibitem{BGNPZ4}
V.~Barone, M.~Genovese,
N.N.~Nikolaev, E.~Predazzi and B.G.~Zakharov,
Phys. Lett. {\bf B328} (1994) 143.


\bibitem{BGNPZ1}
V.~Barone, M.~Genovese,
N.N.~Nikolaev, E.~Predazzi and B.G.~Zakharov,
Phys. Lett. {\bf B268} (1991) 279.

\bibitem{BGNPZ2}
V.~Barone, M.~Genovese,
N.N.~Nikolaev, E.~Predazzi and B.G.~Zakharov,
Phys. Lett. {\bf B304} (1993) 176.


\bibitem{NMC}
P.~Amaudruz {\it et al.}, NMC, Phys. Lett. {\bf B295} (1992) 159.


\bibitem{CCFR3}
S.R.~Mishra {\it et al.} (CCFR), Nevis preprints 1459 and 1465 (1992).

\bibitem{CCFR1}
S.A.~Rabinowitz {\it et al.} (CCFR), Phys. Rev. Lett. {\bf 70}
(1993) 134.

\bibitem{GR}
M.~Gl{\"u}ck, E.~Hoffmann and
E.~Reya, Z. Phys. {\bf C13} (1982) 119;
M.~Gl{\"u}ck, R.M.~Godbole and
E.~Reya, Z. Phys. {\bf C38} (1988) 441;
M.~Gl{\"u}ck, E.~Reya and M.~Stratmann, Nucl. Phys. {\bf B422}
(1994) 37.


\bibitem{CCFR2}
A.O.~Bazarko {\it et al.} (CCFR), Z. Phys. {\bf C65} (1995) 189.

\bibitem{MRS2}
A.D.~Martin, W.J.~Stirling and R.G.~Roberts, Phys. Rev.
 {\bf D50} (1994) 6734.


\bibitem{libri}
E.~Leader and E.~Predazzi,
An Introduction to Gauge Theories
and the New Physics (Cambridge Univ. Press., Cambridge, 1982).\\
R.G.~Roberts, The structure of the proton
(Cambridge Univ. Press., Cambridge, 1990).



%\bibitem{amiconi}
%A.D.Donnachie and P.V.Landshoff,
%{\sl
%DAMTP 93-23} Revised Version, May 93;

\bibitem{Tung}
M.A.G.~Aivazis, J.C.~Collins, F.I.~Olness and W.-K.~Tung, Phys. Rev.
{\bf D50} (1994) 3102.



\bibitem{BDG}
V.~Barone, U.~D'Alesio and M.~Genovese, in preparation.

%\bibitem{nz}
%N.N.Nikolaev and B.G.Zakharov,
%{\sl Z. Phys. } {\bf C49} (1991) 607.


%\bibitem{nos}
%V.Barone, M.Genovese,
%N.N.Nikolaev, E.Predazzi and B.G.Zakharov,
%{\sl Int. J. Mod. Phys.} {\bf A8} (1993) 2779;
%V.Barone, M.Genovese,
%N.N.Nikolaev, E.Predazzi and B.G.Zakharov,
%{\sl Z. Phys. } {\bf C58} (1993) 541.

%\bibitem{mq}
%H.D.~Politzer and H.~Georgi, {\it Phys. Rev. Lett.} {\bf 36} (1976) 1281;
%H.D.~Politzer, {\it Nucl. Phys.} {\bf B117} (1976) 397;
%D.I.~Dyakonov and V.Yu.~Petrov, {\it Nucl. Phys.} {\bf B272} (1986) 457.

%\bibitem{am}
%G.Altarelli and G.Martinelli, {\sl Phys. Lett.} {\bf B76} (1978) 89.

%\bibitem{slow}
%H.~Georgi and H.D.~Politzer, {\it Phys. Rev.} {\bf D14} (1976) 1829;
%R.M.~Barnett, {\it Phys. Rev.} {\bf D14} (1976) 70.


%\bibitem{Brock}
%R.~Brock, {\it Phys. Rev. Lett.} {\bf 44} (1980) 1027.






\end{thebibliography}
\end{document}